\documentclass{iopconfser}
\usepackage{amsmath,cite}

\begin{document}

\title{Midisuperspacetime foam and the cosmological constant}

\author{Steven Carlip$^{1}$}

\affil{$^1$Department of Physics and Astronomy, UC Davis, Davis, CA, USA}
 
\email{carlip@physics.ucdavis.edu}

\begin{abstract}
 Standard quantum field theory arguments predict an enormous cosmological constant.  But what 
 would this mean observationally?  For a homogeneous universe the answer is clear, but if
 the universe is inhomogeneous at the Planck scale, the question becomes more subtle: for a 
 large class of initial data, rapidly expanding and contracting regions coexist and give an average
 expansion near zero.  Classically, such data develop singularities, and we need a quantum 
 description of their evolution.  I describe results from a spherically symmetric midisuperspace 
 model, in which the wave function can become ``trapped'' for long periods in regions in which 
 the average expansion remains small, effectively hiding a large cosmological constant.
 \end{abstract}
 
\section{The cosmological constant problem \label{s1}}
\medskip
 Perhaps the most fundamental principle of general relativity is universality, the fact that gravity
couples with equal strength to every form of energy.  For most cases, this principle has 
been tested to exquisite accuracy \cite{POE}.  But ``every form'' presumably includes the energy of 
the quantum vacuum, which should manifest as a cosmological constant $\Lambda$.  We do not
have an exact computation of this quantity, and standard effective field theory results 
depend sensitively on assumptions about physics beyond the Standard Model.  But even in the
best case, the predictions are at least $55$---and perhaps as many as $122$---orders of magnitude 
larger than what we seem to observe, with a sign determined by the unknown balance of (negative) fermionic 
and (positive) bosonic contributions \cite{Martin}.  This discrepancy, already noted by Pauli in the 
1920s \cite{Strautmann}, has led to desperate measures.  Here I will summarize one such measure, 
introduced in \cite{Carlip1,Carlip2}: the idea that $\Lambda$ could be hidden by Wheeler's ``spacetime 
foam'' \cite{Wheeler}.

Start with a naive question.  If our universe had a huge cosmological constant, how would we know?  
The naive answer is simple: we would see an enormous and rapidly accelerating expansion.  This isn't 
quite right, though.  First, the Einstein field equations are time reversal invariant, so for any fixed $\Lambda$
we could see either expansion or contraction.  More subtly,  we could see expansion \emph{and} contraction.   
Thanks to the beautiful work of Chrusciel, Isenberg, and Pollack \cite{CIP,CIP2}, we know how to construct 
an infinite collection of initial data made up of arbitrarily many expanding an contracting regions, 
glued together by wormholes.  This particular construction is probably not generic, but if we believe
quantum gravity induces Planck scale fluctuations of geometry and that there is no preferred 
microscopic direction of time, we might expect a typical spacetime to consist of a random mixture of 
tiny expanding and contracting regions, effectively hiding any macroscopic sign of a cosmological constant 
\cite{Unruh,Carlip1}.  This would be a realization of ``spacetime foam,'' first proposed by Wheeler in 
part as an attempt to resolve the ultraviolet divergences of quantum field theory \cite{Wheeler}.  Note
that this argument is based on ``foamy'' geometry for a \emph{fixed} $\Lambda$;
fluctuations in vacuum energy might also be important \cite{Unruh,Unruh2}, but are distinct.

 Unfortunately, this does not yet solve the problem.  The classical initial data sets of \cite{CIP,CIP2} quickly 
 develop singularities \cite{Burkhart}, and do not lead to good classical cosmologies.  But it is widely
 expected that quantum gravity will resolve many classical singularities, so it is there we should look.
 
 Again, unfortunately, we are far from having a complete quantum theory of gravity, certainly not one that
 can deal with such complicated geometries.  
 There is, however, a setting that captures  many of the  qualitative features of these ``foamy'' spacetimes 
 while being simple enough to quantize.  This is locally spherically symmetric midisuperspace \cite{Carlip2,Carlip2b}.

\section{Spherically symmetric midisuperspace}
\medskip
Consider the collection of spacetimes, for simplicity with $\Lambda>0$, with metrics that are locally spherically 
symmetric.  By Birkhoff's theorem, any such spacetime looks \emph{locally} like a region of Schwarzschild-de Sitter
space.  But as Morrow-Jones, Witt, and Schleich have shown \cite{MJW,Schleich,Schleich2}, one can glue together 
such regions to form spacetimes that are geometrically and topologically very rich.  For simplicity, let us focus on
the spatial topology $S^2\times S^1$, for which the metric takes the form
\begin{align}
ds^2 = h^2d\psi^2 + f^2(d\theta^2 + \sin^2\theta\, d\varphi^2)
\label{a1}
\end{align}
where $f$ and $h$ are arbitrary periodic functions of the ``radial'' coordinate $\psi$ (as well as $t$).  Topologically one 
can view this manifold as a solid ball with a smaller ball cut out from the center with the inner and outer boundaries 
identified.  Geometrically,  the manifold admits an onion-like structure of concentric shells, joined by necks---essentially 
wormholes---in which the extrinsic curvature goes through zero at the throat.  (Be careful not to picture
these geometries as embedded in flat space.  In particular, the circumferences of successive layers need not be 
monotonic in $\psi$; in some sense we have a ``string of beads'' joined by narrow necks.)

Given such a metric, the momentum and Hamiltonian constraints of general relativity determine the extrinsic curvature 
in terms of $f$ and $h$.  But they determine it only up to a sign, which can change from layer to layer \cite{MJW,Carlip2}.  
The initial data is thus not just the spatial metric, but also a sign for each layer.  While these manifolds differ 
technically from those in \cite{CIP,CIP2}, they are very similar in spirit.

These spacetimes constitute a midisuperspace, a space of metrics for which most of the degrees of freedom 
have been frozen out but infinitely many remain.  This reduction gives us a model that is simple enough to
canonically quantize.  We must, of course, first deal with the notorious ``problem of time'' \cite{Kuchar},
the problem of how to extract time evolution from a frozen formalism.  Common choices for a time variable such 
as volume  and mean curvature fail when one has a mixture of expanding and contracting regions, but the
``dust time'' of Brown and Kucha{\v r} \cite{Brown,Husain}, in which one uses proper time $T$ along a congruence 
of noninteracting ``clocks,'' can be implemented.  The Wheeler-DeWitt equation then becomes \cite{Carlip2}
\begin{align}
i\frac{d\ }{dT}&\Psi[h,f;T] 
    =   \left\{ \frac{3\kappa^2}{64\pi^2}\left(\frac{1}{f^2}\frac{\delta\ }{\delta h}\right)^2 + \frac{\Lambda}{\kappa^2}
   +\frac{3}{\kappa^2f^2}\left(\frac{f^{\prime2}}{h^2}-1\right) + \frac{3\gamma}{f^3}\right\}\Psi[h,f;T] 
   \label{a2}
\end{align}
where $\kappa^2=8\pi G$ and $\gamma$ in an integration constant.  The wave function now has a standard Born 
rule interpretation, with $|\Psi[h,f;T]|^2$ giving a probability density for finding a spatial metric with a given $f$ and 
$h$ at time $T$, though one must be careful to properly gauge fix the inner product \cite{Carlip2,Woodard}.  We 
must also impose the momentum constraints, which now reduce to a single equation
\begin{align}
\frac{i}{8\pi}\left(h\partial_\psi\frac{\delta\ }{\delta h} - f'\frac{\delta\ }{\delta f}\right) \Psi[h,f;T] =0 .
\label{a2a}
\end{align}

Let us now look for solutions of (\ref{a2}--\ref{a2a}) of the form
\begin{align}
\Psi[h,f;T]  = A[h,f]e^{iS[h,f]}e^{-iET}
\label{a3}
\end{align}
in the WKB approximation. The first order approximation determines the phase,
\begin{align}
S[h,f] = &\frac{8\pi}{\kappa^2}\int\!d\psi\,\sigma[h,f;\psi]ff'\left\{\sqrt{1+\beta h^2} - \tanh^{-1}\sqrt{1+\beta h^2} \right\} 
\label{a4}\\[.5ex]
  &\hbox{with}\quad \beta = \frac{f^2}{f^{\prime2}}\left(\frac{\tilde\Lambda}{3} - \frac{1}{f^2} + \frac{\kappa^2\gamma}{f^3}\right)
\quad\hbox{and}\quad
\left\{ \begin{array}{ll}\sigma^2 = 1 \qquad&\hbox{almost everywhere}\\[.5ex]
  \partial_\psi\sigma =0 &\hbox{unless}\  1+\beta h^2 = 0
   \end{array}\right. \nonumber
\end{align}
where back-reaction of the dust shifts the cosmological constant to
$\tilde\Lambda = \Lambda-\kappa^2E$.  The sign $\sigma$ precisely reflects the onion-like structure of the classical
solutions: it is fixed in regions of nonzero extrinsic curvature, but can change at the boundaries where the extrinsic
curvature goes to zero, and its value determines whether a given layer is expanding or contracting.  Given a set of
configurations with $N$ layers, each choice $\{\sigma_1,\dots,\sigma_N\}$ determines a different wave function.
 
The next order WKB approximation determines the amplitude,
 \begin{align}
A = A_0 \exp\left\{\frac{\alpha}{2}\int\!d\psi\,\frac{1}{\sqrt{\beta}}\tan^{-1}(\sqrt{\beta}h)\right\} 
\label{a5}
\end{align}
where $\alpha$ is a constant coming from heat kernel regularization \cite{Carlip2,heat}.  Qualitatively,
this amplitude receives large contributions from the regions around necks joining layers, while thick de Sitter-like layers of 
large expansion are suppressed.   While the inner product is not yet fully understood \cite{Carlip2}, it seems that many-layered
configurations are preferred.   Recall also that each set of signs $\{\sigma_1,\dots,\sigma_N\}$ determines 
a new wave function, so the phase space for ``foamy'' configurations is very large.

\section{Hiding the cosmological constant}
\medskip
The considerations of the preceding section suggest that ``foamy'' configurations, with a mixture of expanding and contracting
regions, dominate.  Our reason for looking at midisuperspace, though, was to try to understand the dynamics.  The use
of a WKB approximation may seem to defeat this purpose, since the wave functions (\ref{a3}) are stationary.  But
recall from introductory quantum mechanics that the WKB approximation \emph{does} contain dynamical information,
in the form of a probability current.

The Wheeler-DeWitt equation (\ref{a2}) has a standard Schr{\"o}dinger form, and admits a probability current
\begin{align}
J[h,f;\psi] = \frac{3i\kappa^2}{64\pi^2}\left(\Psi^*\frac{1}{f^2}\frac{\delta\Psi}{\delta h} - \Psi\frac{1}{f^2}\frac{\delta\Psi^*}{\delta h}\right) .
\label{a6}
\end{align}
This quantity is not diffeomorphism invariant, but one can define invariant modes---see \cite{Carlip2} for details---which
essentially give probability currents among Fourier modes of the metric.  The zero mode, in particular, describes the
probability flow between spaces of different total volumes.  For a large de Sitter-like region, one finds
\begin{align}
J_0 \sim -\frac{3V}{4\pi}\sigma\left(\frac{\Lambda}{3}\right)^{1/2} ,
\label{a7}
\end{align}
which for large $\Lambda$ describes a rapid flow in volume, with a direction determined by the sign $\sigma$.  
For ``foamy'' configurations, on the other hand, one must sum an expression similar to (\ref{a7}) over many layers with 
random signs, which will tend to cancel.  Probabilities can thus become trapped, flowing out of regions of spacetime 
foam only very slowly.  Hence the picture described in \S\ref{s1}, in which a large cosmological 
constant might be hidden at macroscopic scales, seems to persist dynamically in a quantum model.

What can we conclude from this model?  To start with, although spacetime foam is a plausible idea \cite{Carlip3}, we 
don't know whether it is actually an outcome of the full quantum theory of gravity.  Even if it is, we don't know whether 
this simple midisuperspace model is sophisticated enough to capture the relevant behavior.  Nevertheless, we have a start.  
In a model that plausibly has some of the key features of  full quantum gravity, we have seen that the observational consequences
of a very large cosmological constant can be suppressed.  ``Foamy'' regions of midisuperspace, in which the macroscopic
average expansion is small, have relatively high probabilities, and a wave function concentrated in such regions can be
trapped there by quantum fluctuations.  

There is still much to understand, even in this rather simple model.  The full inner product is not yet known (although
it is known that it will not alter the cancellations in the probability current \cite{Carlip2}).  The structure of the
midisuperspace  needs further analysis; it would be good to be able to better demarcate regions of midisuperspace and
the flows of probability among them.  We don't know enough good diffeomorphism-invariant observables, and
we have not yet constructed time-dependent wave packets.  And it would be useful to see this problem from 
a completely different point of view, perhaps using a path integral to study the nucleation of
contracting bubbles in an expanding spacetime.

Let me conclude with a new and even more speculative idea.  If a large cosmological constant is hidden by cancellation between
expanding and contracting regions, one would not expect this cancellation to be exact, but only to hold up to a
factor of order $1/\sqrt{N}$, where $N$ is the relevant number of fluctuating regions.  As Sorkin pointed out in a slightly 
different context \cite{Sorkin}, the volume of our past light cone is approximately $10^{240}\ell_P{}^4$, giving
$1/\sqrt{N}\sim 10^{-120}$.  In the present context, this coincidence should probably not be taken too seriously:
what is averaged here is the expansion, which goes as $\sqrt{\Lambda}$, so the effective cosmological constant
is $\Lambda_{\mathit{eff}}\sim \Lambda/N$.  If, however, the ultimate quantum theory of gravity is holographic,
it is worth noting that the areas of most candidate ``screens''---the maximum area of our past light cone, our particle horizon,
our apparent horizon---are all of order $10^{120}$ in Planck units, a number that would
reduce a Planck scale cosmological constant to a value  $\Lambda_{\mathit{eff}}$ comparable to the one we observe.

\newpage
\noindent{\bf Acknowledgments}\\
 
 \noindent This work was supported in part by US Department of Energy grant DE-FG02-91ER40674.

\end{document}